\documentclass[twocolumn,prl]{revtex4}
\usepackage{color}
\usepackage{graphicx}
\usepackage{dcolumn}
\usepackage{bm}
\usepackage{hyperref}
\usepackage{amsmath}
\usepackage{latexsym}
\usepackage{amssymb}
\usepackage{layout}
\usepackage{verbatim}
\usepackage{amsfonts,epsfig}
\newcommand{\ket}[1]{\left|#1\right>}
\newcommand{\bra}[1]{\left< #1 \right|}
\newcommand{\beq}{\begin{equation}}
\newcommand{\eeq}{\end{equation}}
\newcommand{\bea}{\begin{eqnarray}}
\newcommand{\eea}{\end{eqnarray}}
\newcommand{\nn}{\nonumber}
\newcommand{\tr}{\hbox{Tr}}

\newcommand{\mean}[1]{\langle{#1}\rangle{}}

\newcommand{\os}{\chi}

\begin{document}

\author{Edwin Barnes$^{1}$\email{barnes@umd.edu}, {\L}ukasz Cywi{\'n}ski$^2$, and S. Das Sarma$^{1}$}
\affiliation{ $^{1}$Condensed Matter Theory Center, Department of
Physics, University of Maryland, College Park, Maryland 20742-4111, USA\\
$^{2}$Institute of Physics, Polish Academy of Sciences, al.~Lotnik{\'o}w 32/46, PL 02-668 Warszawa, Poland}

\title{Nonperturbative master equation solution of central spin dephasing dynamics}

\begin{abstract}
We solve the long-standing central spin problem for a general set of inhomogeneous bath couplings and a large class of initial bath states.
We compute the time evolution of the coherence of a central spin coupled to a spin bath by resumming all orders of the
time-convolutionless master equation, thus avoiding the need to assume weak coupling to the bath. The fully quantum,
non-Markovian solution is obtained in the large-bath limit and is valid up to a timescale set by the largest coupling constant. Our result captures the full decoherence
of an electron spin qubit coupled to a nuclear spin bath in a GaAs quantum dot for experimentally relevant parameters. In addition, our solution is quite compact and can readily be used to make quantitative predictions for the decoherence process
and to guide the design of nuclear state preparation protocols.
\end{abstract}

\maketitle
Since the central spin model was first introduced by Gaudin \cite{Gaudin} several decades ago, it has appeared in
diverse physical settings
such as spin dynamics in disordered insulators \cite{Sachdev_JAP87}, interacting electrons in
metallic grains \cite{Amico_PRL01}, and more recently in semiconductor spin qubits \cite{Khaetskii_PRB03,Dobrovitski_PRE03,Coish_PRB04,Yao_PRB06,Bortz_PRB07,Cywinski_PRL09,Cywinski_PRB09,Coish_PRB10,Erbe_PRL10,Barnes_PRB11}.
It has also been widely studied in the
context of integrable systems, where close connections to BCS theory and related theories of pairing forces have been discovered \cite{Amico_PRL01,Yuzbashyan_JPA05}.
In addition, several variants of the central spin model have served as toy models for comparing and contrasting various master equation formalisms \cite{Breuer_PRB04,Fischer_PRA07,Ferraro_PRB08}.

This broad interest in the central spin model has led to a plethora of disparate approaches to computing its dynamics. Since its inception, it has been recognized as
an integrable system which admits a solution via Bethe ansatz \cite{Gaudin,Amico_PRL01,Bortz_PRB07}.
However, this yields only a very complicated ground state wavefunction, and progress in extracting dynamical information for the central spin has been limited; to date only modest bath sizes of up to 30 bath spins have been treated in this manner \cite{Bortz_PRB07}. Sophisticated numerical recipes for calculating the central spin evolution have been developed \cite{Dobrovitski_PRE03,Al_Hassanieh_PRL06,Zhang_JPC07}, but success tends to require either small baths (tens of spins) or the assumption that the initial bath state is completely unpolarized. These requirements are too restrictive for many applications in the context of III-V semiconductor spin qubits, where the number of nuclear spins ranges from $10^4$ to $10^6$, and polarized baths are employed to facilitate qubit operations and extend coherence times \cite{Foletti_NP09,Bluhm_PRL10}.

This state of affairs led to the development of several approximate analytical methods to compute the central spin evolution. Some of these methods employ an effective pure-dephasing Hamiltonian which is derived perturbatively from the central spin Hamiltonian using a canonical transformation \cite{Yao_PRB06,Cywinski_PRL09,Cywinski_PRB09}. This approach culminated with a nonperturbative solution of the effective Hamiltonian dynamics \cite{Cywinski_PRL09,Cywinski_PRB09} and has been successful in describing spin echo experiments in the case of unpolarized baths \cite{Bluhm_NP11}. However, an uncontrolled approximation in the effective-Hamiltonian derivation makes it unclear when this approach is valid, and it has yet to be extended to more general bath states. Methods employing generalized master equations, on the other hand, offer a controlled approximation and naturally describe polarized bath states, but only perturbative treatments have been given so far \cite{Coish_PRB04,Fischer_PRA07,Ferraro_PRB08,Coish_PRB10,Barnes_PRB11}, leading to solutions which are valid only outside the regime
relevant for many semiconductor spin qubit experiments \cite{Petta_Science05,Koppens_PRL08,Barthel_PRL09,Bluhm_NP11}.

In this Letter, we solve the central spin problem using the time-convolutionless (TCL) master equation for a general set of inhomogeneous coupling constants and a large set of initial bath states, including both polarized and unpolarized baths. The TCL equation is an exact equation for the reduced density matrix of a system coupled to a bath; although this equation is time-local, it incorporates the full bath dynamics \cite{breuer,Fischer_PRA07,Ferraro_PRB08}. With only a very modest condition on the magnetic field, we give a closed-form solution describing the evolution up to a timescale set by the largest bath coupling. We are therefore presenting an exact solution to the central spin model as it pertains to gated GaAs spin qubits since this temporal window contains the entire decay of the electron spin coherence in the low magnetic field regime where the central spin model gives a good description of the physics \cite{footnote1}. Although we will focus on the example of spin qubits, our results are potentially applicable to any Gaudin-type central spin problem. This result is also important to the general study of open quantum systems; we are not aware of other examples involving a large, nontrivial and highly non-Markovian bath where an all-orders resummation of a master equation expansion is performed.

The central spin model is comprised of a central spin coupled to a spin bath via a Heisenberg interaction. Assuming a nonzero external magnetic field, the Hamiltonian is
\beq
\hat{H} = \sum_{k}A_{k} \mathbf{S}\cdot \mathbf{I}_{k} + \Omega S^{z} + \sum_{k} \omega_{k} I^{z}_{k} \,\, ,  \label{ham}
\eeq
with $\mathbf{S}$ denoting the central spin operator,  $\mathbf{I}_{k}$ the bath spins, $A_{k}$ the (hyperfine) couplings, and $\Omega$ and $\omega_{k}$ the central spin and bath spin Zeeman energies, respectively. In the context of semiconductor electron spin qubits, the $A_k$ are determined by the shape of the electron wavefunction envelope, but we will leave the $A_{k}$ completely arbitrary. We refer to $\sum_k A_k S^zI_k^z$ as the Overhauser term, and $V_{\text{ff}} \! = \! {1\over2}\sum_k A_k(S^+I_k^-+S^-I_k^+)$ as the flip-flop term, where $S^{\pm}=S^x\pm i S^y$, and similarly for $I_k^\pm$.
The total interaction energy is  $\mathcal{A} \! \equiv \! \sum_{k}A_{k}$, and the number of bath spins appreciably interacting with the central spin is $N\equiv{\cal A}^2/(\sum_kA_k^2)$. Roughly speaking, the bath produces an effective (Overhauser) magnetic field, the magnitude of which is controlled by $\cal A$ and the bath polarization, and about which the central spin precesses, while the $A_k$ set the scale for the precession of individual bath spins about the central spin.

The TCL equation is an exact equation for the time evolution of the reduced density matrix of a system coupled to a bath \cite{breuer}. Although this equation contains full memory of the bath dynamics, unlike equations such as the Nakajima-Zwanzig equation \cite{Coish_PRB04,Coish_PRB10,Barnes_PRB11}, it has the attractive feature that it is a time-local ordinary differential equation. Working in an interaction picture defined with respect to $\hat H_0=\sum_{k}A_{k} S^zI_{k}^z + \Omega S^{z} + \sum_{k} \omega_{k} I^{z}_{k}$ and denoting the total density matrix in the interaction picture by $\rho(t)$, the TCL equation has the form
\beq
{d\over dt}P\rho(t)=\sum_{n=1}^\infty{\cal K}_n(t)P\rho(t).\label{TCLgen}
\eeq
The operator $P$ projects the full density matrix onto the reduced density matrix of the system; its precise definition will be given shortly.
Eq. (\ref{TCLgen}) is defined in terms of a perturbative expansion in $V_{\text{ff}}$ which is nominally controlled by the quantity ${\cal A}/\Omega$. Ultimately our solution will not require this expansion to be convergent since we will sum the entire series, so ${\cal A}/\Omega$ need not be small. The $n$th-order kernel ${\cal K}_n(t)$ encapsulates full bath effects arising from $n$th-order flip-flop processes in which the central spin flips $n$ times with one or more bath spins. It can be expressed as an integral of ordered cumulants involving $P$ and the interaction-picture Liouville operator $L$ (defined by $\dot\rho=-i L\rho$); for example the second-order kernel is ${\cal K}_2(t)=-\int_0^t dt' PL(t)L(t')P$. The rules for constructing the higher order kernels can be found in Ref.~\cite{breuer}.

We will assume that the initial density matrix separates into system and bath components, $\rho(0)=\rho_S(0)\otimes\rho_B(0)$, and we will use a set of ``correlated projectors" \cite{Fischer_PRA07}, in which case the action of $P$ on a matrix $M$ is $PM \! = \!\sum_\alpha \tr_B\{\Pi_\alpha M\}\otimes{1\over{\cal N}_\alpha}\Pi_\alpha$,
where the $\Pi_\alpha$ are a set of bath projectors satisfying $\Pi_\alpha\Pi_\beta \! = \!\delta_{\alpha\beta}\Pi_\beta$ and $\sum_\alpha\Pi_\alpha \! = \!1$,
and ${\cal N}_\alpha=\tr_B\{\Pi_\alpha\}$. The $\Pi_\alpha$ allow us to write the reduced density matrix of the system as a sum of independent degrees of freedom: $\rho_S \! = \! \sum_\alpha \tr_B\{\Pi_\alpha\rho\}\equiv \sum_\alpha\rho_S^{(\alpha)}$.
The form of the TCL equation we use requires $P\rho(0)=\rho(0)$, which in turn implies that the choice of the $\Pi_\alpha$ will constrain the possible initial bath states. An appropriate choice of the $\Pi_\alpha$ can either simplify or vastly improve the convergence of the TCL equation depending on the symmetries of the Hamiltonian and initial bath state \cite{Fischer_PRA07,Barnes_PRB11}; for now we leave the $\Pi_\alpha$ completely general.

Since we want to compute the off-diagonal component of $\rho_S(t)$ (coherence function), we multiply Eq. (\ref{TCLgen}) by $S^+$ and trace over both system and bath to obtain
\beq
\dot\rho_{S,-+}(t)=\tr_S\{S^+\dot\rho_S(t)\}=\sum_{n=1}^\infty\tr\{S^+{\cal K}_n(t)P\rho(t)\}.\label{TCLgen2}
\eeq
The summand on the right-hand side of this equation is comprised of integrals of terms with the general structure $\tr\{S^+{\cal L}_1P{\cal L}_2P\ldots {\cal L}_r P\rho(t)\}$, where ${\cal L}_i$ represents a string of Liouville operators $L(t_{i_1})L(t_{i_2})\ldots$. If one assumes that $\Pi_\alpha$ is such that bath correlators of the type $\tr_B\{\Pi_\alpha I_{\ell_1}^\pm I_{\ell_2}^\pm\ldots\}$ vanish unless they contain equal numbers of raising and lowering operators, then it is straightforward to show \cite{supplement} that in the case of the central spin model (\ref{ham}), such terms factorize:
\bea
&&\tr\{S^+{\cal L}_1P{\cal L}_2P\ldots {\cal L}_r P\rho(t)\}\nn\\&&=\sum_\alpha {1\over{\cal N}_\alpha^r}\rho_{S,-+}^{(\alpha)}(t)\prod_{i=1}^r\tr\{S^+{\cal L}_iS^-\Pi_\alpha\}.\label{factorize}
\eea
Since every term on the right-hand side of Eq. (\ref{TCLgen2}) factorizes in this way, we can expand the left-hand side as $\dot\rho_{S,-+}(t)=\sum_\alpha\dot\rho_{S,-+}^{(\alpha)}(t)$ and separately equate each term of the $\alpha$-sum.
The resulting set of equations is readily solved:
\beq
\rho_{S,-+}^{(\alpha)}(t)=\rho_{S,-+}^{(\alpha)}(0)\exp\left\{\sum_n{\cal G}_n^{(\alpha)}(t)\right\},
\eeq
where ${\cal G}_n^{(\alpha)}(t)\equiv\int_0^t dt'\tr\{S^+{\cal K}_n(t')S^-\Pi_\alpha\}.$

To calculate ${\cal G}_n^{(\alpha)}(t)$ we must first compute correlators of the type $\tr\{S^+{\cal L}S^-\Pi_\alpha\}$ where ${\cal L}$ is an arbitrarily long string of Liouville operators. Restricting ourselves for simplicity to the case $\omega_k=\omega$, we find \cite{supplement}
\bea
&& \!\!\!\!\! \tr\{S^+ L(t_{b_1})
\ldots L(t_{b_{2q}})S^-\Pi_\alpha\}
 \approx \frac{1}{4^q} e^{-i\Omega_\alpha\sum_{i=1}^{2q}(-1)^{b_i}t_{b_i}} \nn \\ &&
\times
 \sum_{k=0}^{q}\binom{q}{k}\tr_B\{(h^+h^-)^{k}\Pi_\alpha(h^-h^+)^{q-k}\}.\label{approximations}
\eea
We have defined the operators $h^j\equiv \sum_\ell A_\ell I_\ell^j$ and frequency $\Omega_\alpha\equiv \Omega-\omega+\tr_B\{h^z\Pi_\alpha\}/{\cal N}_\alpha$, and we have assumed that $[h^z,\Pi_\alpha]=0$. The latter condition ensures that the correlator vanishes for odd numbers of Liouville operators, which in turn implies that ${\cal G}_n^{(\alpha)}$ vanishes for odd $n$. There are two approximations being made in Eq. (\ref{approximations}). The first approximation assumes $t\ll1/A_{max}$, where $A_{max} \! \sim \! \mathcal{A}/N$ is the largest coupling, and this leads to the time-dependence appearing only as a phase factor in Eq. (\ref{approximations}). For spin qubits in GaAs with $N=10^6$, $1/A_{max}$ can be on the order of $10\mu$s, long enough to capture the full decay of the electron spin coherence for experimentally relevant values of the magnetic field \cite{Petta_Science05,Koppens_PRL08,Barthel_PRL09,Bluhm_NP11}. For comparison, we can also consider spin qubits in Si \cite{Maune_Nature12}, in which case $1/A_{max}$ is on the order of $250\mu$s.

If this were the only approximation, then the right-hand side of Eq. (\ref{approximations}) would have an additional sum over permutations of the $t_{b_i}$, but only a certain subset of these permutations were kept in Eq. (\ref{approximations}). Retaining only this subset amounts to keeping the leading order terms in the $\Omega_\alpha t\gg1$ limit at each order of the TCL expansion. To illustrate the nature of this RPA-like approximation, we consider its effect on the lowest-order terms. If we kept all permutations, then the second-order term would have the form ${\cal G}_2^{(\alpha)}(t)\sim \delta_\alpha^2u(t,\Omega_\alpha)$ while the fourth-order terms stemming from the $q=2$, $k=0,2$ cases in (\ref{approximations}) would have the form ${\cal G}_4^{(\alpha)}(t)\sim \delta_\alpha^4v(t,\Omega_\alpha)$ where $\delta_\alpha\equiv {{\cal A}\over\sqrt{N}\Omega_\alpha}$ and $u(t,\Omega_\alpha)=i\Omega_\alpha t+e^{-i\Omega_\alpha t}-1$, $v(t,\Omega_\alpha)=-\Omega_\alpha^2t^2-4i\Omega_\alpha t-(6+2i\Omega_\alpha t)e^{-i\Omega_\alpha t}+6$. The RPA-like approximation amounts to taking $u\to i\Omega_\alpha t$ and $v\to -\Omega_\alpha^2t^2$. Self-consistency of the approximation requires $\delta_\alpha\ll 1$, which imposes a lower bound on the magnetic field (corresponding to a few mT for gated dots in GaAs). The necessity for $\delta_\alpha\ll1$ can be seen from $u$ and $v$ by noting that we can only neglect the linear (in $t$) term in $v$ if it is small compared to the linear term in $u$. Later on, we will see that this approximation captures the envelope of the coherence function, and we will also find that we can relax this approximation to a large degree by keeping the full form of ${\cal G}_2^{(\alpha)}(t)$.

With Eq. (\ref{approximations}) in hand, it is straightforward to assemble these correlators into the function ${\cal G}_n^{(\alpha)}(t)$ using the rules for constructing the TCL ordered cumulants \cite{supplement}:
\begin{widetext}
\beq
{\cal G}_n^{(\alpha)}(t)=\left(it\over4\Omega_\alpha\right)^{n/2}
\sum_{\{q_i\}\in{\cal P}(n/2)}{1\over\prod_{i=1}^r q_i!}{(-1)^{r+1}\over r{\cal N}_\alpha^r}\prod_{i=1}^r\sum_{k=0}^{q_i}\binom{q_i}{k}
\tr_B\{(h^+h^-)^{k}\Pi_\alpha(h^-h^+)^{q_i-k}\}.\label{Gn}
\eeq
\end{widetext}
In this expression, $\{q_i\}\in{\cal P}(n/2)$ means that $\{q_i\}$ is an ordered integer partition of $n/2$, with $r$ being the number of $q_i$ comprising the partition \cite{footnote2}. To evaluate this expression, we make perhaps the simplest choice for the projectors: $\Pi_\alpha=\Pi_\os=\ket{\os}\bra{\os}$, where $\ket{\os}\equiv \bigotimes_k \ket{I_k,m_k^\os}$ is a product of eigenstates of the $I_k^z$ ($I_k(I_k+1)$ and $m_k^\os$ are eigenvalues of $\mathbf{I}_k^2$ and $I_k^z$). In this case, ${\cal N}_\alpha=1$. This particular choice is well suited to applications pertaining to spin qubits \cite{Coish_PRB04,Coish_PRB10,Barnes_PRB11}.
With an initial bath density matrix of the form $\rho_B(0)=\sum_\os \rho_{\os\os}\ket{\os}\bra{\os}$, it is possible to perform the various sums in (\ref{Gn}) despite their complexity \cite{supplement}, and we find to leading order in the limit of large $N$ that the coherence function $W(t)\! \equiv \! \tilde\rho_{S,-+}(t)/\tilde\rho_{S,-+}(0)$ in the Schr\"odinger picture is
\bea
W(t)=\sum_\os {\rho_{\os\os}(d_\os^+-d_\os^-)e^{i(\Omega + h^z_\os)t}\over d_\os^+e^{-{it\over4\Omega_\os}(d_\os^+-d_\os^-)}-d_\os^-e^{{it\over4\Omega_\os}(d_\os^+-d_\os^-)}},\label{coherence}
\eea
where $h^z_\os\equiv\tr_B\{\Pi_\chi h^z\}$ is the Overhauser field associated with the state $\ket{\os}$, $d_\os^\pm\equiv\bra{\os}h^\mp h^\pm\ket{\os}= \sum_\ell A_\ell^2[I_\ell(I_\ell+1)-m_\ell^\os(m_\ell^\os\pm1)]$ quantifies transverse fluctuations of this field,
 and $\Omega_\os\equiv \Omega-\omega+h^z_\os$ is the difference between the effective Zeeman energy of the central spin ($\Omega+h^z_\os$) and the bath spin Zeeman energy ($\omega$).
Equation (\ref{coherence}) is the main result of this paper; it describes the envelope of the coherence function of the central spin for an arbitrary set of couplings $A_k$ and for a large set of initial bath states up to time $t\lesssim 1/A_{max}$.

\begin{figure}
\begin{center}
\includegraphics[height=3cm, width=.48\columnwidth]{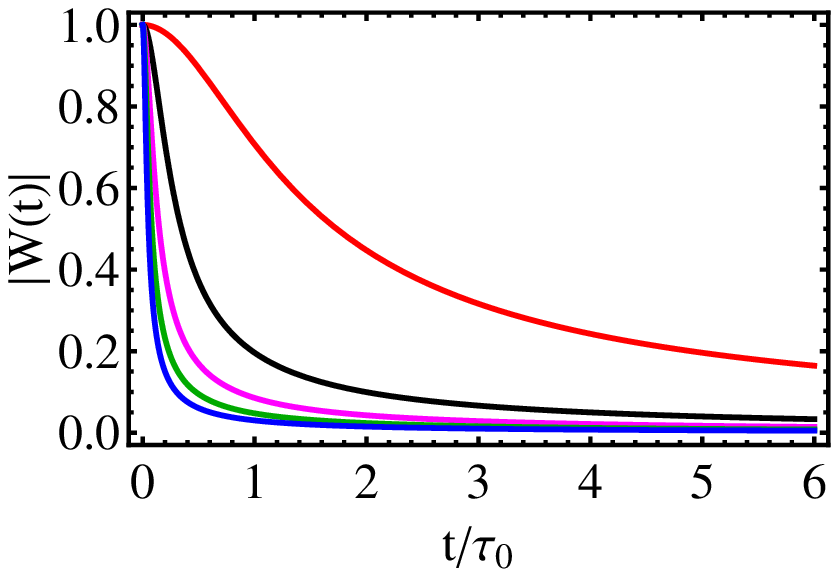}
\includegraphics[height=3cm, width=.48\columnwidth]{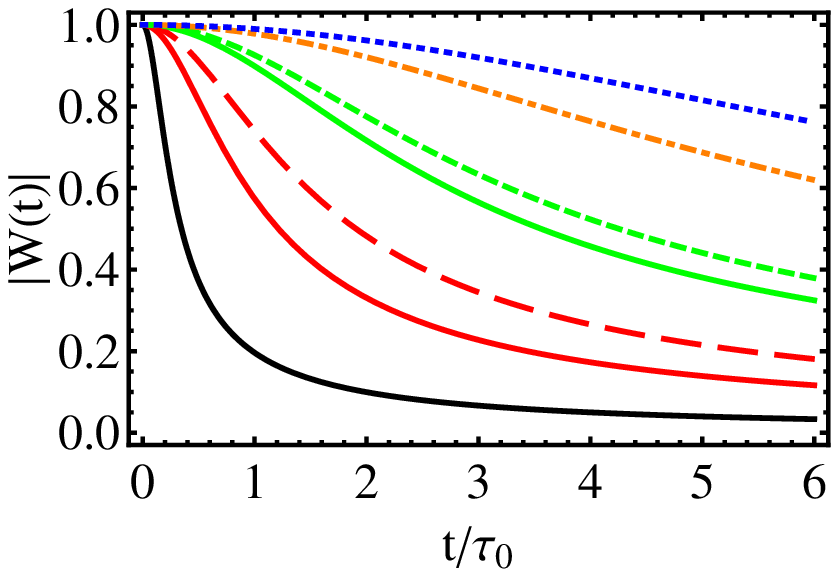}
\caption{\label{fig:varying_I_and_p} (Color online) Coherence function (Eq. (\ref{uniform})). Left panel: changing bath total spin $I$ with $p=0$ and $p^2_\perp={2\over3}(I+1)$. From top to bottom, $I$ takes the values ${1\over2}, {3\over2}, {5\over2}, {7\over2}, {9\over2}$. Right panel: changing bath polarization $p$ with $I={3\over2}$, ${\cal A}/\Omega=30$, and $p^2_\perp={5\over3}$. From top to bottom, $p$ takes the values ${3\over4}, {1\over2}, {1\over4}, -{1\over4}, {1\over10}, -{1\over10}, 0$.}
\end{center}
\end{figure}

As a first example, we consider the case of a uniformly polarized bath \cite{footnote3} with all bath spins having the same total angular momentum $I$, and for which $h^z_\os$ is the same for all $\os$ \cite{Coish_PRB04,Coish_PRB10}. Writing $h^z_\os={\cal A}Ip$ and $\Omega_p \! = \! \Omega-\omega+{\cal A}Ip$,
where $p={1\over NI}\sum_\os\rho_{\os\os}\sum_km_k^\os$ is the average
polarization of the bath, with $p=0$ denoting an unpolarized bath and $|p|=1$ maximal polarization, we find
\bea
W(t)={pe^{i(\Omega+{\cal A}Ip)t}\over p\cos\left({2Ipt\over \tau_p}\right)
-ip^{2}_{\perp}\sin\left({2Ipt\over \tau_p}\right)},\label{uniform}
\eea
with $p_\perp^2\equiv I+1-{1\over NI}\sum_\os\rho_{\os\os}\sum_k(m_k^\os)^2$
(for $I\!= \!1/2$, $p^2_\perp=1$), and $\tau_p\equiv4N\Omega_p/{\cal A}^2$. Setting $p{=}0$ in (\ref{uniform}) yields the zero-polarization result, $W(t)=1/(1-2iIp^2_\perp t/\tau_p)$, obtained in previous works \cite{Cywinski_PRL09,Cywinski_PRB09,Barnes_PRB11} using less rigorous methods. The left panel of Fig.~\ref{fig:varying_I_and_p} shows that the decoherence rate increases with increasing $I$ due to a corresponding increase in the number of bath degrees of freedom. For electron spin qubits in a GaAs nuclear spin bath ($I=3/2$) where the magnetic field is typically on the order of 100 mT \cite{Petta_Science05} so that ${\cal A}/\Omega\approx 30$ and $\tau_0\approx (2/15)N/{\cal A}$, it is clear that the coherence decays almost completely before time $t=1/A_{max}\approx N/{\cal A}$ is reached. The right panel of Fig.~\ref{fig:varying_I_and_p} depicts the extent to which the decoherence time increases with increasing bath polarization, a well known effect which can be understood in terms of a reduction of phase space for flip-flops. It is also evident that positive net polarization leads to longer decoherence times compared with negative polarization, since for the latter the decrease of $|\Omega_{p}|$ facilitates virtual flip-flops. Our results provide quantitative predictions for the enhancement of decoherence time resulting from bath polarization.

\begin{figure}[htp]
  \includegraphics[height=3cm, width=.48\columnwidth]{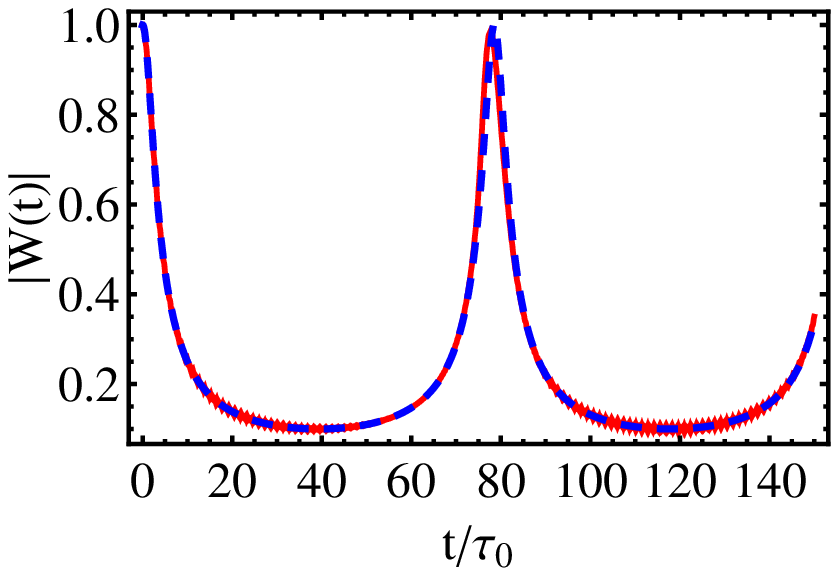}
  \includegraphics[height=3cm, width=.48\columnwidth]{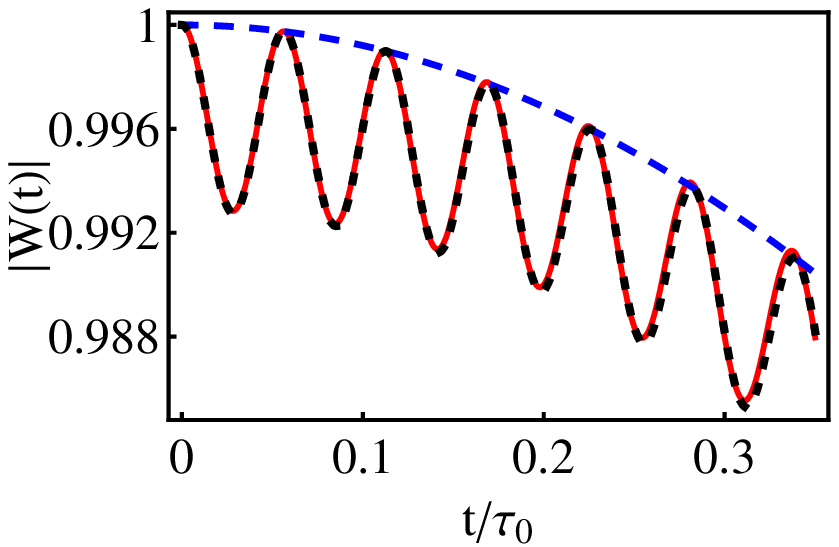}
  \caption{(Color online) Left panel: Coherence function from Eq. (\ref{uniform}) (blue, dashed) vs. exact box model solution from \cite{Barnes_PRB11} (red, solid) with $I={1\over2}$, ${\cal A}/\Omega=30$, $N=10^4$, $p={1\over10}$. Right panel: Zoom-in of left panel with modified coherence function $\widetilde W(t)$ (black, dotted) included as well.}\label{fig:tcl_vs_box}
\end{figure}

A salient feature of the uniform-polarization result, Eq. (\ref{uniform}), is that it depends on the couplings only through the quantities ${\cal A}$ and $N$. This implies that any set of couplings which yield the same values of ${\cal A}$ and $N$ will give rise to the same central spin evolution for $t\ll N/{\cal A}$. For example, this evolution should be reproduced by a model in which all the couplings are equal, $A_k={\cal A}/N$, the so-called ``box" model, which is exactly solvable (see e.g. \cite{Barnes_PRB11} for the solution in the case of a polarized bath). A comparison of Eq. (\ref{uniform}) with the exact box model solution is shown in the left panel of Fig.~\ref{fig:tcl_vs_box}, and it is evident that the two solutions agree very well. This insensitivity to the particular values of the $A_k$ on timescales $t\ll N/{\cal A}$ was anticipated in \cite{Barnes_PRB11} based on energy-time uncertainty; here, we have given a direct proof of this result, and we show below that its validity requires uniform polarization. The exact box model solution is not known to have a closed form, so the fact that Eq. (\ref{uniform}) constitutes a very good approximate closed-form solution is an added bonus of the present work (Eq. (\ref{coherence})).
\begin{figure}[htp]
  \includegraphics[height=3cm, width=.48\columnwidth]{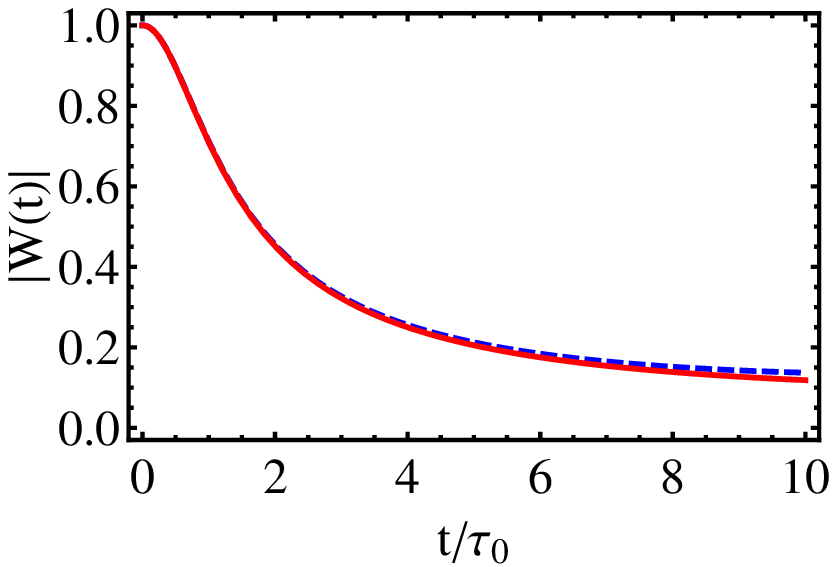}
  \includegraphics[height=3cm, width=.48\columnwidth]{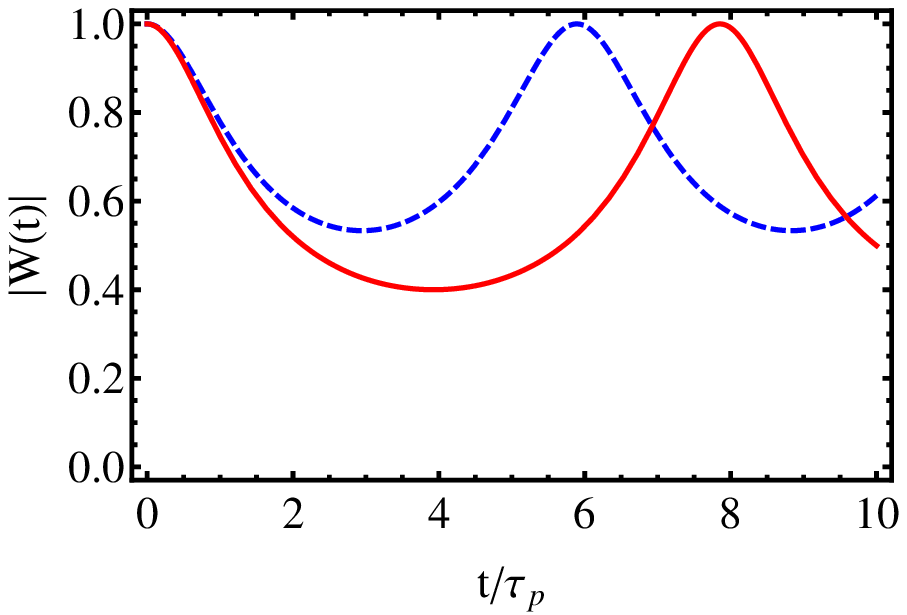}
  \caption{(Color online) Coherence function with uniform polarization Eq. (\ref{uniform}) (red, solid) vs. non-uniform polarization Eq. (\ref{nonuniform}) (blue, dashed) with $I={1\over2}$, ${\cal A}/\Omega=30$, and (left panel) $p={1\over10}$, $\tau_p={1\over3}{N\over{\cal A}}$, (right panel) $p={2\over5}$, $\tau_p={14\over15}{N\over{\cal A}}$.}\label{fig:uniform_vs_non}
\end{figure}

The right panel of Fig.~\ref{fig:tcl_vs_box} reveals that Eq. (\ref{uniform}) does not capture a high-frequency, small-amplitude modulation exhibited by the exact box model solution \cite{Barnes_PRB11}. However, it turns out that it is easy to correct for this by relaxing the RPA-like approximation at second order of the TCL flip-flop expansion, in which case one finds that ${\cal G}_2^{(\os)}={1\over4\Omega_\os^2}[i\Omega_\os t+e^{-i\Omega_\os t}-1](d^+_\os+d^-_\os)$, leading to the modified coherence function $\widetilde W=W\exp[(e^{-i\Omega_p t}-1)/(\Omega_p\tau_p)]$. This modified function reproduces very well the small modulations as is evident in the right panel of Fig.~\ref{fig:tcl_vs_box}. The condition $\delta_\os={{\cal A}\over \sqrt{N}\Omega_\os}\ll1$ ensures that higher-order corrections to these modulations are negligible.

We stress that the coherence function given in Eq. (\ref{coherence}) can describe more general bath polarization states. To illustrate this, we consider a simple example of a non-uniformly polarized initial bath state where the bath spins possess an average polarization which depends linearly on their coupling: $\mean{m_k}=\sum_\os\rho_{\os\os}m_k^\os=NIpA_k/{\cal A}$. This describes a bath configuration where the spins closest to the central spin are more polarized, while those further away are less polarized. In the case of spin qubits in quantum dots, this is qualitatively a physically plausible configuration since the nuclear spins are typically polarized through manipulation of the central electron spin \cite{Foletti_NP09,Bluhm_PRL10,Barnes_PRL11,Latta_NP09,Vink_NP09,Gullans_PRL10}. This time, the coherence function computed from Eq. (\ref{coherence}) depends not only on $\sum_kA_k$ and $\sum_kA_k^2$, but also on $\sum_kA_k^3$, meaning that the result now depends on details of the distribution of the $A_{k}$. Assuming the $A_k$ are distributed in accordance with a two-dimensional Gaussian wavefunction of the central spin electron, and taking the continuum limit (e.g.~$\sum_{k}f(A_{k}) \rightarrow \int_{0}^{A_{max}} \rho(A) f(A) dA$ with $\rho(A) \! = \! {N\over2A}$ and $A_{max} \! = \! {2\mathcal{A}\over N}$), we obtain
\bea
W(t)={pe^{i(\Omega+{\cal A}Ip)t}\over p\cos\left({8Ipt\over 3 \tau_p}\right)
-{3i\over4} p^{2}_{\perp}\sin\left({8Ipt\over 3\tau_p}\right)}.\label{nonuniform}
\eea
Fig.~\ref{fig:uniform_vs_non} shows that while for lower bath polarizations this result is similar to what we found in the uniform-polarization case, for larger polarizations, the differences between the two solutions become quite pronounced. This illustrates how our results can be used to distinguish between different narrowed polarization distributions produced using empirical nuclear state preparation schemes \cite{Latta_NP09,Vink_NP09,Foletti_NP09,Bluhm_PRL10} which are not yet understood microscopically.

In conclusion, we have presented a non-perturbative solution to the central spin problem. Our solution is an elegant, closed-form expression which is valid for an arbitrary set of couplings to the spin bath and for a large class of initial bath states. In the context of GaAs spin qubits, our result solves completely the pure-hyperfine decoherence problem and can be used to quantitatively predict decoherence times and to better understand nuclear state preparation protocols.

E.B. would like to thank S. Economou for useful discussions. This work is supported by LPS-NSA and IARPA. {\L}C acknowledges support from the Homing programme of the Foundation for Polish Science supported by the EEA Financial Mechanism.

\newpage

\begin{widetext}

\section{Supplementary information}

This supplement shows explicitly the intermediate algebraic steps which arise at different stages of the derivation of the central spin coherence function.

\section{Factorization of bath correlators}

In the main text, it is shown that the derivative of the coherence function in the interaction picture can be expressed as
\beq
\dot\rho_{S,-+}(t)=\tr_S\{S^+\dot\rho_S(t)\}=\sum_{n=1}^\infty\tr\{S^+{\cal K}_n(t)P\rho(t)\},\label{TCLgen2}
\eeq
where ${\cal K}_n(t)$ is the $n$th order TCL kernel. This quantity is given in terms of ordered cumulants involving the Liouville operator $L(t)$ and the projection operator $P$. For example, at fourth order we have
\bea
{\cal K}_4(t)&=&\int_0^tdt_1\int_0^{t_1}dt_2\int_0^{t_2}dt_3\Big[PL(t)L(t_1)L(t_2)L(t_3)P-PL(t)L(t_1)PL(t_2)L(t_3)P\nn\\&&-PL(t)L(t_2)PL(t_1)L(t_3)P -PL(t)L(t_3)PL(t_1)L(t_2)P\Big].\label{4thordertclkernel}
\eea
The rules for constructing these kernels can be found in Ref.~\cite{breuersupp}. For the central spin Hamiltonian, the explicit action of the Liouville operator on a matrix $\rho$ is
\beq
L\left(\begin{matrix} \rho_{++}& \rho_{+-} \cr \rho_{-+}&\rho_{--}\end{matrix}\right)={1\over2}\left(\begin{matrix} h_{+-}\rho_{-+}-\rho_{+-}h_{-+} \,\,\, & h_{+-}\rho_{--}-\rho_{++}h_{+-}\cr h_{-+}\rho_{++}-\rho_{--}h_{-+} \,\,\, & h_{-+}\rho_{+-}-\rho_{-+}h_{+-}\end{matrix}\right),\label{liouville}
\eeq
with
\bea
h_{+-}&\equiv&e^{i\Omega t}\sum_\ell A_\ell e^{-i(\omega_\ell-A_\ell/2)t}e^{ih^zt}I_\ell^-,\nn\\
h_{-+}&\equiv&e^{-i\Omega t}\sum_\ell A_\ell e^{i(\omega_\ell+A_\ell/2)t}e^{-ih^zt}I_\ell^+=h_{+-}^\dagger.\label{defofh}
\eea
The summand on the right-hand side of Eq.~(\ref{TCLgen2}) is comprised of integrals of terms with the general structure $\tr\{S^+{\cal L}_1P{\cal L}_2P\ldots {\cal L}_r P\rho(t)\}$, where ${\cal L}_i$ represents a string of Liouville operators $L(t_{i_1})L(t_{i_2})\ldots$. We will now show how such terms factorize by focusing on the case $r=2$. Because of the particular form of the central spin Hamiltonian, when we expand $\tr\left\{S^+{\cal L}_1P{\cal L}_2P\rho(t)\right\}$ using Eqs.~(\ref{liouville}) and (\ref{defofh}), each resulting term depends on either $(P{\cal L}_2P\rho)_{-+}$ or $(P{\cal L}_2P\rho)_{+-}$. Terms depending on $(P{\cal L}_2P\rho)_{-+}$ always contain equal numbers of bath raising and lowering operators, while terms depending on $(P{\cal L}_2P\rho)_{+-}$ contain unequal numbers. If we assume $\Pi_\alpha$ is such that $\tr_B\{\Pi_\alpha I_{\ell_1}^\pm I_{\ell_2}^\pm\ldots\}$ vanishes for unequal numbers of $I_{\ell}^+$ and $I_{\ell}^-$, then only the terms with $(P{\cal L}_2P\rho)_{-+}$ are non-vanishing, and we have
\bea
\tr\left\{S^+{\cal L}_1P{\cal L}_2P\rho(t)\right\}&=&\tr\{S^+{\cal L}_1(P{\cal L}_2P\rho)_{-+}S^-\}=\tr\{S^+{\cal L}_1\left(\sum_\alpha\tr_B\{\Pi_\alpha {\cal L}_2P\rho\}\otimes{1\over{\cal N}_\alpha}\Pi_\alpha\right)_{-+}S^-\}\nn\\
&=&\sum_\alpha{1\over{\cal N}_\alpha}\tr\{S^+{\cal L}_1S^-\Pi_\alpha\}\tr\{S^+{\cal L}_2P\rho\Pi_\alpha\}
=\sum_\alpha{1\over{\cal N}_\alpha^2}\rho_{S,-+}^{(\alpha)}\tr\{S^+{\cal L}_1S^-\Pi_\alpha\}\tr\{S^+{\cal L}_2S^-\Pi_\alpha\}.\nn\\&&
\eea
This factorization generalizes straightforwardly to correlators with any number of $P$ insertions:
\beq
\tr\{S^+{\cal L}_1P{\cal L}_2P\ldots {\cal L}_r P\rho(t)\}=\sum_\alpha {1\over{\cal N}_\alpha^r}\rho_{S,-+}^{(\alpha)}(t)\prod_{i=1}^r\tr\{S^+{\cal L}_iS^-\Pi_\alpha\}.\label{factorize}
\eeq
The assumption that $\tr_B\{\Pi_\alpha I_{\ell_1}^\pm I_{\ell_2}^\pm\ldots\}$ vanishes for unequal numbers of raising and lowering operators is an assumption that restricts the possible choices we make for $\Pi_\alpha$, and thus the possible choices of initial bath state. More specifically, if we represent $\Pi_\alpha$ as a matrix in the basis of eigenstates of $h^z$, then this restriction forces many of the off-diagonal elements to vanish. This restriction is rather mild and still leaves us with a large class of initial bath states including those relevant in the case of electron spin decoherence in the presence of a polarized nuclear spin bath.

\section{Evaluation of bath correlators}

The factorization shown in Eq.~(\ref{factorize}) allows us to reduce the problem of computing correlators of the form $\tr\{S^+{\cal L}_1P{\cal L}_2P\ldots {\cal L}_r P\rho(t)\}$ to computing correlators of the form $\tr\{S^+{\cal L}S^-\Pi_\alpha\}$. The latter can be evaluated straightforwardly:
\beq
{1\over{\cal N}_\alpha}\tr\left\{S^+L(t')L(t_1)\ldots L(t_{2q-1})S^-\Pi_\alpha\right\}=
{1\over4^q}\sum_{k=0}^q \sum_{\{a_i\}} g_k^{(\alpha)}(\{t_{a_i}\}),\label{corrworho}
\eeq
with
\beq
g_k^{(\alpha)}(\{t_{a_i}\})={1\over{\cal N}_\alpha}\tr_B\left\{h_{-+}(t_{a_1})h_{+-}(t_{a_2})\ldots h_{-+}(t_{a_{2k-1}})h_{+-}(t_{a_{2k}})\Pi_\alpha h_{+-}(t_{a_{2k+1}})h_{-+}(t_{a_{2k+2}})\ldots h_{+-}(t_{a_{2q-1}})h_{-+}(t_{a_{2q}})\right\}.
\eeq
Here, the sum $\sum_{\{a_i\}}$ means that we sum over all permutations of the $a_i$, where $i=1\ldots 2q$ and $a_i\in[0,q-1]$, such that $a_1<a_2<\ldots<a_{2k}$ and $a_{2k+1}>a_{2k+2}>\ldots>a_{2q}$. We define $t_0\equiv t'$. At this point, we further restrict $\Pi_\alpha$ to be such that $[\Pi_\alpha,h^z]=0$, and we focus on the short-time evolution, $t<1/A_{max}$, in which case we may neglect the $A_k$ appearing in the exponents in the expressions for $h_{+-}$ and $h_{-+}$ in Eq.~(\ref{defofh}). In this case, $g_k^{(\alpha)}$ simplifies to
\beq
g_k^{(\alpha)}(\{t_{a_i}\})={1\over{\cal N}_\alpha}e^{-i\Omega_\alpha(t_{a_1}-t_{a_2}+\ldots-t_{a_{2k}}-t_{a_{2k+1}}+t_{a_{2k+2}}-\ldots+t_{a_{2q}})}\tr_B\left\{(h^+h^-)^{k}\Pi_\alpha(h^-h^+)^{q-k} \right\},
\eeq
where $h^\pm\equiv \sum_\ell A_\ell I_\ell^\pm$.

The second (RPA-like) approximation we will make is to keep only those permutations of the $a_i$ which yield the dominant terms at order $n$ after the integrations over the $t_{a_i}$ are performed. Retaining only this subset of permutations amounts to keeping the leading order terms in the $\Omega_\alpha t\gg1$ limit at each order of the TCL expansion. For example, one such dominant permutation is given by $t_{a_1}=t'$, $t_{a_i}=t_{i-1}$, $t_{a_j}=t_{2q+2k-j}$, $i=2\ldots 2k$, $j=2k+1\ldots 2q$, which leads to the time-dependent factor
\beq
e^{-i\Omega_\alpha\sum_{i=1}^{2q}(-1)^{a_i}t_{a_i}}.\label{timedep}
\eeq
In fact, the nested time integrals (see for instance Eq.~(\ref{4thordertclkernel})) are such that the time-dependent factors in the dominant (in the $\Omega_\alpha t\gg1$ limit) terms of $\tr\{S^+{\cal L}S^-\Pi_\alpha\}$ always have the form shown in Eq.~(\ref{timedep}). This means that we need to sum over only those permutations of the $a_i$ which preserve Eq.~(\ref{timedep}). There are $\binom{2q}{2k}$ different choices of the $a_i$, and $\binom{q}{k}$ of these will lead to the same time dependence as in Eq.~(\ref{timedep}). This counting can be understood by noticing that the quantity $t-t_1+t_2-\ldots-t_{2k-1}-t_{2k}+\ldots+t_{2q}$ is invariant under pairwise swaps of $t_{2i}-t_{2i+1}$ with $t_{2j}-t_{2j+1}$. All such swaps which preserve the condition that the first $2k$ times in this quantity are monotonically increasing from left to right and the last $2q-2k$ times are monotonically decreasing from left to right should be counted as separate contributions to the time-dependent factor. Restricting ourselves for simplicity to the case $\omega_k=\omega$, we therefore have
\beq
\tr\{S^+ L(t_{a_1})L(t_{a_2})\ldots L(t_{a_{2q}})S^-\Pi_\alpha\}
\approx{1\over4^q}e^{-i\Omega_\alpha\sum_{i=1}^{2q}(-1)^{a_i}t_{a_i}}\sum_{k=0}^{q}\binom{q}{k}\tr_B\{(h^+h^-)^{k}\Pi_\alpha(h^-h^+)^{q-k}\}.\label{approximations}
\eeq
As explained in the main text, this RPA-like approximation is controlled by the small quantity $\delta_\alpha\equiv {{\cal A}\over\sqrt{N}\Omega_\alpha}$, which places a lower limit on the external magnetic field.

Now that we have obtained Eq.~(\ref{approximations}), we need to combine these correlators according to Eq.~(\ref{factorize}) and integrate the result over the $t_i$. Since the time-dependent factors combine the same way for every term at order $n$, these integrations are easily performed all at once:
\beq
\int_0^t dt'\int_0^{t'}dt_1\int_0^{t_1}dt_2\ldots\int_0^{t_{n-2}}dt_{n-1}e^{-i\Omega_\alpha(t'-t_{1}+\ldots+t_{n-2}-t_{n-1})} ={(-i)^{n/2}\over(n/2)!}\left(t\over\Omega_\alpha\right)^{n/2}.
\eeq
Defining ${\cal G}_n^{(\alpha)}(t)\equiv\int_0^t dt'\tr\{S^+{\cal K}_n(t')S^-\Pi_\alpha\}$, we then find
\beq
{\cal G}_n^{(\alpha)}(t)=\left(it\over4\Omega_\alpha\right)^{n/2}
\sum_{\{q_i\}\in{\cal P}(n/2)}{1\over\prod_{i=1}^r q_i!}{(-1)^{r+1}\over r{\cal N}_\alpha^r}\prod_{i=1}^r\sum_{k=0}^{q_i}\binom{q_i}{k}
\tr_B\{(h^+h^-)^{k}\Pi_\alpha(h^-h^+)^{q_i-k}\},\label{Gn}
\eeq
where $\{q_i\}\in{\cal P}(n/2)$ means that $\{q_i\}$ is an ordered integer partition of $n/2$, with $r$ being the number of $q_i$ comprising the partition. For example, if $n=6$, then the possible choices of the $q_i$ are $\{q_i\}=\{3\},\{2,1\},\{1,2\},\{1,1,1\}$, and $r$ takes the values $1,2,2,3$ respectively. Summing over these partitions and including the combinatoric factor $(n/2)!/( r\prod_{i=1}^r q_i!)$ is tantamount to summing over the different terms appearing in the $n$th order TCL kernel (as dictated by the rules for constructing ordered cumulants \cite{breuersupp}), for example the four terms in Eq.~(\ref{4thordertclkernel}). An additional factor of $(-1)^{n/2+r+1}$ also comes directly from the rules for constructing ${\cal K}_n(t)$.

\section{Coherence function for a certain class of initial bath states}

Eq.~(\ref{Gn}) determines the coherence function through the relation
\beq
\rho_{S,-+}(t)=\sum_\alpha\rho_{S,-+}^{(\alpha)}(t)=\sum_\alpha\rho_{S,-+}^{(\alpha)}(0)\exp\left\{\sum_n{\cal G}_n^{(\alpha)}(t)\right\}.\label{coherence1}
\eeq
We will now evaluate this for a certain class of initial bath states which is relevant for the electron spin decoherence problem. These states have the form $\rho_B(0)=\sum_\os \rho_{\os\os}\ket{\os}\bra{\os}$ (which corresponds to choosing $\Pi_\alpha=\ket{\os}\bra{\os}$), where $\ket{\os}\equiv \bigotimes_k \ket{I_k,m_k^\os}$ is a product of eigenstates of the $I_k^z$ ($I_k(I_k+1)$ and $m_k^\os$ are eigenvalues of $\mathbf{I}_k^2$ and $I_k^z$). In this case, ${\cal N}_\alpha=1$. In other words, we are focusing on initial bath states which are mixtures of states possessing well defined Overhauser fields.

The first step in evaluating Eq.~(\ref{coherence1}) is to evaluate the correlators in Eq.~(\ref{Gn}),
\beq
\tr_B\{(h^+h^-)^{k}\Pi_\alpha(h^-h^+)^{q_i-k}\}=\bra{\os}(h^-h^+)^{q_i-k}(h^+h^-)^{k}\ket{\os}.
\eeq
For fixed values of $q_i$ and $k$, this correlator is the sum of all possible ways of contracting raising operators $I_{\ell_1}^+$ with lowering operators $I_{\ell_2}^-$. In general, a single contraction can involve an arbitrary number of operators (each sharing the same nuclear site index $\ell$), but in the large $N$ limit, one can ignore all contractions involving more than two operators since these are far fewer in number relative to pairwise contractions and should therefore have a subleading effect. This is tantamount to keeping only those contributions in which each raising and lowering operator pair has a site index which is distinct from every other pair. For example, one fourth-order contribution comes from the correlator
\beq
\bra{\os}h^-h^+h^-h^+\ket{\os}=\sum_{\{\ell_j\}}\left(\prod_jA_{\ell_j}\right)\bra{\os}I_{\ell_1}^-I_{\ell_2}^+I_{\ell_3}^-I_{\ell_4}^+\ket{\os}.
\eeq
This correlator has two distinct pairwise contractions: $\ell_1=\ell_2$, $\ell_3=\ell_4$ and $\ell_1=\ell_4$, $\ell_2=\ell_3$. The first contraction gives the contribution
\beq
\sum_{\ell_1\ne\ell_3}A_{\ell_1}^2A_{\ell_3}^2\bra{\os}I_{\ell_1}^-I_{\ell_1}^+\ket{\os}\bra{\os}I_{\ell_3}^-I_{\ell_3}^+\ket{\os}
=\sum_{\ell_1\ne\ell_3}A_{\ell_1}^2A_{\ell_3}^2c^{(\os)+}_{\ell_1}c^{(\os)+}_{\ell_3},
\eeq
where we have defined $c^{(\os)\pm}_{\ell}\equiv\bra{\os}I_{\ell}^\mp I_{\ell}^\pm\ket{\os}$, while the second contraction gives
\beq
\sum_{\ell_1\ne\ell_2}A_{\ell_1}^2A_{\ell_2}^2\bra{\os}I_{\ell_1}^-I_{\ell_1}^+\ket{\os}\bra{\os}I_{\ell_2}^+I_{\ell_2}^-\ket{\os}
=\sum_{\ell_1\ne\ell_2}A_{\ell_1}^2A_{\ell_2}^2c^{(\os)+}_{\ell_1}c^{(\os)-}_{\ell_2}.
\eeq
Therefore, the net result in the large $N$ limit is
\beq
\bra{\os}h^-h^+h^-h^+\ket{\os}=\sum_{\ell_1\ne\ell_2}A_{\ell_1}^2A_{\ell_2}^2[c^{(\os)+}_{\ell_1}c^{(\os)+}_{\ell_2}+c^{(\os)+}_{\ell_1}c^{(\os)-}_{\ell_2}].
\eeq
To simplify this slightly further, we will invoke the large $N$ limit once again to eliminate the inequality condition between $\ell_1$ and $\ell_2$. Then defining
\beq
d_\os^\pm\equiv \sum_{\ell}A_{\ell}^2c^{(\os)\pm}_{\ell}=\sum_{\ell}A_{\ell}^2[I_\ell(I_\ell+1)-m_\ell^\os(m_\ell^\os\pm1)],
\eeq
we may write
\beq
\bra{\os}h^-h^+h^-h^+\ket{\os}=d_\os^+(d_\os^++d_\os^-).
\eeq
A correlator involving $q_i$ pairs of raising and lowering operators will have $q_i!$ different pairwise contractions which will each contribute to the coherence function. Each contraction will contribute a monomial involving $d_\os^+$ and $d_\os^-$, so that the full correlator will contribute some polynomial in these quantities. For example, the correlator with $q_i=3$ and $k=2$ involves six contractions and evaluates to
\beq
\bra{\os}h^-h^+h^+h^-h^+h^-\ket{\os}=2(d_\os^+)^2d_\os^-+4d_\os^+(d_\os^-)^2.
\eeq
When all of the correlators contributing to the coherence at a given order $n$ are added together, the result is again a polynomial in $d_\os^\pm$. Interestingly, this polynomial (for $n>2$) is related to Eulerian numbers:
\beq
\sum_{\{q_i\}\in{\cal P}(n/2)}{1\over\prod_{i=1}^r q_i!}{(-1)^{r+1}\over r{\cal N}_\alpha^r}\prod_{i=1}^r\sum_{k=0}^{q_i}\binom{q_i}{k}
\tr_B\{(h^+h^-)^{k}\Pi_\alpha(h^-h^+)^{q_i-k}\}={2^{n/2}\over(n/2)!}d_{\os}^+d_{\os}^-\sum_{k=0}^{{n\over2}-2}\left\langle\begin{matrix} {n\over2}-1 \cr k \end{matrix} \right\rangle (d_{\os}^+)^k(d_{\os}^-)^{{n\over2}-k-2}.\label{formula2}
\eeq
The symbol $\left\langle\begin{matrix} p \cr q \end{matrix} \right\rangle$ denotes an Eulerian number, which for positive integer $p$ and non-negative integer $q$ can be expressed as
\beq
\left\langle\begin{matrix} p \cr q \end{matrix} \right\rangle=\sum_{j=0}^{q+1}(-1)^j\binom{p+1}{j}(q-j+1)^p.
\eeq
The sum over $k$ in (\ref{formula2}) can be expressed as a polylogarithm:
\beq
d_{\os}^+d_{\os}^-\sum_{k=0}^{{n\over2}-2}\left\langle\begin{matrix} {n\over2}-1 \cr k \end{matrix} \right\rangle (d_{\os}^+)^k(d_{\os}^-)^{{n\over2}-k-2}
=(d_{\os}^+-d_{\os}^-)^{n/2}\hbox{Li}_{1-{n\over2}}\left({d_{\os}^-\over d_{\os}^+}\right)=(d_{\os}^+-d_{\os}^-)^{n/2}\sum_{j=1}^\infty\left({d_{\os}^-\over d_{\os}^+}\right)^jj^{{n\over2}-1}.
\eeq
The final expression involving the infinite sum is strictly speaking only valid for $d_{\os}^-<d_{\os}^+$ (which is true for negative polarizations $\sum_\ell A_\ell^2m_\ell^\os<0$), but we will tacitly perform an analytic continuation at the end of the calculation. We may then express the $\os$th coherence degree of freedom as
\beq
{\rho_{S,-+}^{(\os)}(t)\over\rho_{S,-+}^{(\os)}(0)}=\exp\bigg[{it\over4\Omega_{\os}}(d_{\os}^++d_{\os}^-)+\sum_{j=1}^\infty\left({d_{\os}^-\over d_{\os}^+}\right)^j{1\over j}\sum_{n>2}\left(itj(d_{\os}^+-d_{\os}^-)\over2\Omega_{\os}\right)^{n/2}{1\over(n/2)!}\bigg].
\eeq
The sum over $n$ is now easily done:
\beq
\sum_{n>2}\left(itj(d_{\os}^+-d_{\os}^-)\over2\Omega_{\os}\right)^{n/2}{1\over(n/2)!}
=\exp\bigg[{itj(d_{\os}^+-d_{\os}^-)\over2\Omega_{\os}}\bigg]-1-{itj(d_{\os}^+-d_{\os}^-)\over2\Omega_{\os}},
\eeq
and the resulting sums over $j$ are readily identified as logarithmic, so that
\beq
\sum_{j=1}^\infty\left({d_{\os}^-\over d_{\os}^+}\right)^j{1\over j}\sum_{n>2}\left(itj(d_{\os}^+-d_{\os}^-)\over2\Omega_{\os}\right)^{n/2}{1\over(n/2)!}
=\log\bigg[{d_{\os}^+-d_{\os}^-\over d_{\os}^+-d_{\os}^-\exp[{it\over2\Omega_{\os}}(d_{\os}^+-d_{\os}^-)]}\bigg]-{it\over2\Omega_{\os}}d_{\os}^-.
\eeq
Adding this result to the second order ($n=2$) contribution and exponentiating, we finally obtain
\beq
{\rho_{S,-+}^{(\os)}(t)\over\rho_{S,-+}^{(\os)}(0)}={d_\os^+-d_\os^-\over d_\os^+e^{-{it\over4\Omega_\os}(d_\os^+-d_\os^-)}-d_\os^-e^{{it\over4\Omega_\os}(d_\os^+-d_\os^-)}}.
\eeq
The full coherence function is obtained by summing over degenerate Overhauser states. To express the coherence function in the Schr\"odinger picture, we simply include the phase factor $e^{i(\Omega+h^z_\os)t}$ inside the sum over $\os$, where $h^z_\os\equiv \tr_B\{\Pi_\os h^z\}$; this follows immediately from the property $[\Pi_\os,\hat H_0]=0$. Denoting the Schr\"odinger-picture density matrix as $\tilde\rho$, the Schr\"odinger-picture coherence function is then
\beq
W(t)\equiv{\tilde\rho_{S,-+}(t)\over\tilde\rho_{S,-+}(0)}=\sum_\os {\rho_{\os\os}(d_\os^+-d_\os^-)e^{i(\Omega+h^z_\os)t}\over d_\os^+e^{-{it\over4\Omega_\os}(d_\os^+-d_\os^-)}-d_\os^-e^{{it\over4\Omega_\os}(d_\os^+-d_\os^-)}}.\label{coherence}
\eeq

\end{widetext}


\begin{thebibliography}{31}
\bibitem{footnote2} For example, if $n=6$, then the possible choices of the $q_i$ are $\{q_i\}=\{3\},\{2,1\},\{1,2\},\{1,1,1\}$, and $r$ takes the values $1,2,2,3$ respectively.
\bibitem{footnote3} By uniform polarization we mean that $\sum_{\os}\rho_{\os\os}m^{\os}_{k}$, $\sum_\os \rho_{\os\os} m_k^\os m_\ell^\os$, etc. are independent of the choice of spin labels $k$, $\ell$, ...
\end{thebibliography}
\end{document}